\documentclass[12pt,epsfig]{article}
\usepackage{graphicx,amsmath,amssymb}

\newlength{\dinwidth}
\newlength{\dinmargin}
\def\lapproxeq{\lower .7ex\hbox{$\;\stackrel{\textstyle
<}{\sim}\;$}}
\def\gapproxeq{\lower .7ex\hbox{$\;\stackrel{\textstyle
>}{\sim}\;$}}
\def\gtrsim{\lower .7ex\hbox{$\;\stackrel{\textstyle
>}{\sim}\;$}}
\def\lesim{\lower .7ex\hbox{$\;\stackrel{\textstyle
<}{\sim}\;$}}

\def\be{\begin{equation}}
\def\ee{\end{equation}}
\def\bea{\begin{eqnarray}}
\def\eea{\end{eqnarray}}

\def\qq{q\bar{q}}

\def\GeV{\rm GeV}

\begin{document}
\begin{flushright}
IPPP/06/36 \\
DCPT/06/72 \\
19th June 2006 \\

\end{flushright}

\vspace*{0.5cm}

\begin{center}
{\Large \bf Information from leading neutrons at HERA}

\vspace*{1cm}
\textsc{V.A.~Khoze$^{a,b}$, A.D. Martin$^a$ and M.G. Ryskin$^{a,b}$} \\

\vspace*{0.5cm}
$^a$ Department of Physics and Institute for
Particle Physics Phenomenology, \\
University of Durham, DH1 3LE, UK \\
$^b$ Petersburg Nuclear Physics Institute, Gatchina,
St.~Petersburg, 188300, Russia \\

\end{center}

\vspace*{0.5cm}

\begin{abstract}
In principle, leading neutrons produced in photoproduction and deep-inelastic scattering at HERA have the potential to determine the pion structure function, the neutron absorptive cross section and the form of the pion flux. To explore this potential we compare theoretical predictions for the $x_L$ and $p_t$ spectra of leading neutrons, and the $Q^2$ dependence of the cross section, with the existing ZEUS data.
\end{abstract}

\section{Introduction}

The original motivation for measuring leading neutrons at HERA was to determine the pion structure function $F_2^{\pi}(x,Q^2)$, assuming that the process $\gamma p \to Xn$ is dominated by $\pi$-exchange and that the $\pi$-flux is known from the analysis of hadronic data in the triple-Regge region.  The problem, however, is that the original $\pi$-flux, given by the simple $\pi$-exchange diagram, is modified by ``soft'' rescattering effects which are different in deep inelastic scattering and hadron-hadron collisions.  In hadron-hadron collisions the leading neutron has a much larger probability of a secondary interaction, which changes the longitudinal fraction, $x_L$, of the proton's momentum carried by the neutron, as well as the neutron's transverse momentum, $p_t$. On the contrary, for deep-inelastic scattering the probability of secondary interactions becomes negligible; and so knowing the $\pi$-flux, we can directly extract $F_2^{\pi}(x,Q^2)$ at large $Q^2$.

At HERA we have the possibility of varying $Q^2$. So the increasing role of rescattering can be studied as we go to lower $Q^2$, and finally to photoproduction.  It was shown in Ref.~\cite{KKMRln}, that for $x_L>0.6$, rescattering may just be considered as {\it absorption} of the fast neutron. After rescattering the neutron {\it migrates} to lower $x_L$, but, at large $x_L$, the population due to migration is small and here rescattering acts as absorption.  Thus, for large $x_L$, absorptive effects may be described in terms of the survival factor, $S^2$, of the rapidity gap associated with $\pi$-exchange.  $S^2$ may be calculated in the usual way (see, for example, \cite{KMRsoft}); its value depends on the inelastic neutron cross section.  Note that a $\gamma ^*$-initiated process proceeds through a $\gamma^* \to \qq$ transition, followed by the interaction of the $\qq$-pair with the target.  The probability of rescattering is therefore determined by the cross section of this $\qq$ interaction.  The size of the $\qq$-pair decreases with increasing $Q^2$, and the effect of rescattering becomes negligible at large $Q^2$. Therefore leading neutron data for a range of $Q^2$ gives information about the $\qq$-neutron cross section, or $\sigma_{\rm abs}$.  The preliminary analysis of the data indicates that both $F^{\pi}_2$ and $\sigma_{\rm abs}$ are in reasonable agreement with theoretical expectations based on the additive quark model.

The next step is the possibility of studying the precise form of the $\pi$-flux by measuring the $p_t$ dependence of the neutron yields.  Thus, thanks to the additional parameter $Q^2$, it seems possible to extract three types of information from one set of leading neutron deep-inelastic data:
\begin{itemize}
\item[(a)]
the pion structure function $F_2^{\pi}(x,Q^2)$,
\item[(b)]
the $(\qq ) -N$ absorptive cross section $\sigma_{\rm abs}$, where $N=n$ or $p$,
\item[(c)]
the form of the $\pi$-flux; the normalization of the $\pi$-flux is given by the known $G_{\pi NN}$ coupling constant, when the pion is close to its mass shell.
\end{itemize}
Unfortunately, at present, the $p_t$ dependence measured at HERA is flatter than that expected from theoretical models---that is from any reasonable parametrization of the pion form factor.

In Section 2 we compare our cross section predictions for leading neutrons, as a function of $x_L$, with the HERA data, and in Section 3 we study the $p_t$ dependence.  In order to account for the flatter $p_t$ dependence of the data, in Section 4, we study the possible role of $\rho$ and $a_2$ exchange contributions. We consider both photoproduction and deep-inelastic production of leading neutrons.

\section{Photo- and DIS-production as a function of $x_L$}
\begin{figure}
\begin{center}
\includegraphics[height=5cm]{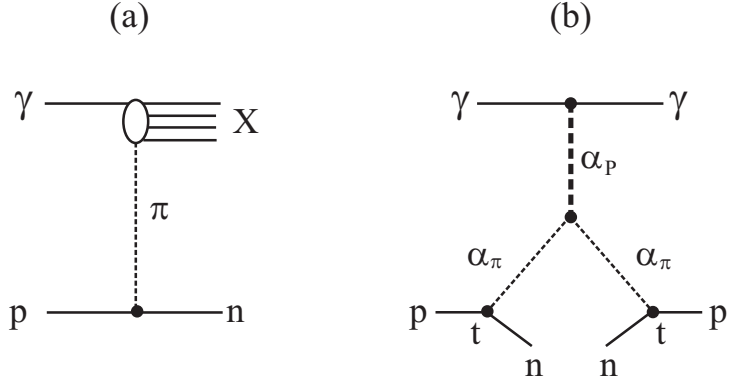}
\caption{(a) The pion-exchange amplitude and (b) the corresponding dominant triple-Regge contribution to the
cross section of the inclusive production of leading
neutrons at HERA, $\gamma p \to Xn$.
\label{fig:1}}
\end{center}
\end{figure}

If we assume $\pi$-exchange dominance, then the inclusive production of leading neutrons, $\gamma p \to Xn$ of Fig.~\ref{fig:1}(a), is given by the triple-Regge diagram shown in Fig.~\ref{fig:1}(b). We have
\be
\frac{d\sigma(\gamma p \to Xn)}{dx_L dt}~=~S^2~\frac{G_{\pi^+ pn}^2}{16\pi^2}~\frac{~(-t)}{(t-m^2_{\pi})^2}~F^2(t)~(1-x_L)^{1-2\alpha_\pi (t)}~\sigma^{\rm tot}_{\gamma\pi}(M^2),
\label{pi}
\ee
where the coefficient of $\sigma^{\rm tot}_{\gamma\pi}$ is called the pion flux. The pion trajectory, $\alpha_\pi(t)=\alpha^\prime_\pi (t-m^2_{\pi})$, is taken to have slope $\alpha^\prime_\pi \simeq
1~\GeV ^{-2}$, and the $\pi^+ pn$ coupling constant is $G^2_{\pi^+ pn}/8\pi
=13.75$ \cite{GpiN}.  The invariant mass $M$ of the produced system $X$
is given by $M^2 \simeq s(1-x_L)$. $F(t)$ is the form factor resulting from the pion-nucleon and $\pi\pi P$ vertices with off-mass-shell pions, see Fig.~\ref{fig:1}(b).  The survival factor $S^2$, which takes into account absorptive corrections, depends on $x_L$ and $p_t$ of the leading neutron. The calculation of $S^2$ is outlined in the Appendix. 

\begin{figure}
\begin{center}
\includegraphics[height=15cm]{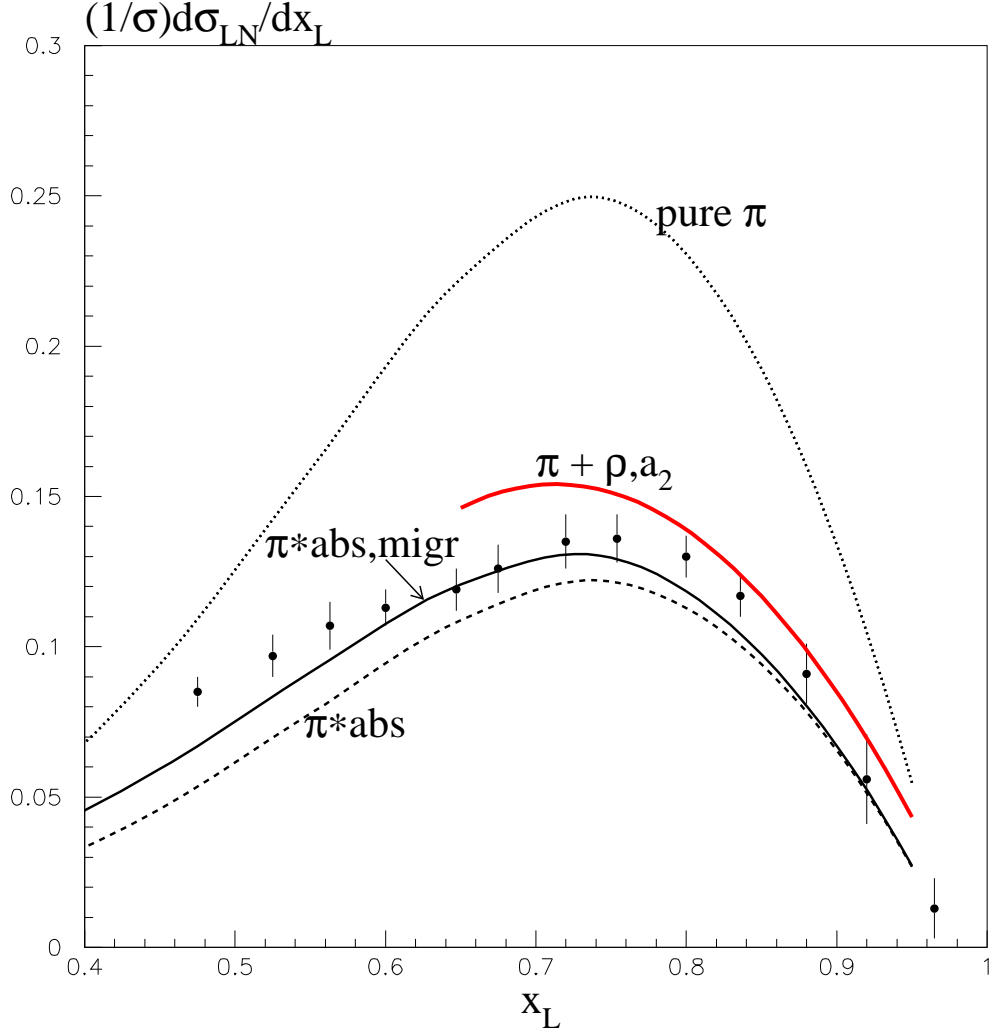}
\caption{The predictions for the $x_L$ spectra of photoproduced leading neutrons compared with preliminary ZEUS data \cite{dis06}; only the systematic errors on the data points are indicated, as these dominate the statistical errors.  The dotted, dashed and lower continuous
curves are respectively the results assuming first only Reggeised $\pi$ exchange, then including
absorptive effects, and finally allowing for migration; the calculation is described in \cite{KKMRln}, updated here to allow for the different experimental cuts. The upper continuous curve corresponds to including $\rho$ and $a_2$ exchange contributions, as well as $\pi$ exchange, as described in Section 4.
\label{fig:2}}
\end{center}
\end{figure}

The cross section of the $\gamma \pi$-interaction, $\sigma^{\rm tot}_{\gamma\pi}$, and the pion structure function, $F_2^{\pi}$, are the quantities measured in photoproduction and deep inelastic scattering respectively, where
\be
\sigma^{\rm tot}_{\gamma^*\pi}~=~\frac{4\pi^2 \alpha}{Q^2}~F_2^\pi.
\label{sigF}
\ee
We use the additive quark model to obtain theoretical estimates, assuming for photoproduction
\be
\sigma^{\rm tot}_{\gamma \pi}~=~\frac{2}{3}\sigma^{\rm tot}_{\gamma p},
\label{aqmph}
\ee
and for deep-inelastic scattering\footnote{Unfortunately, the present parametrizations of the parton distributions of the pion are unreliable in the low $x$ region of interest. Therefore we take (\ref{DIS}).} 
\be
F_2^\pi (x,Q^2)~=~\frac{2}{3}~F_2^p (\frac{2}{3}x,Q^2).
\label{DIS}
\ee
We rescale the Bjorken $x$ in order to have the same energy for the $\gamma^*$-valence $q$ interaction. Another possibility, which we will discuss, is to simultaneously rescale $Q$ by the ratio of the pion and proton radii. It was shown in Ref.~\cite{KKMRln}, that if we take a reasonable value of the neutron absorption cross section\footnote{The value taken was motivated by the $\rho$-dominance model of the photon.} then this approach satisfactorily describes the ZEUS data for the photoproduction of leading neutrons at large $x_L$.  The description, updated for the new experimental cuts used in \cite{dis06}, is shown in Fig.~\ref{fig:2}. From the figure we see that the absorptive corrections reduce the cross section, given simply by Reggeised pion
exchange, by a factor $S^2$, averaged over $p_t^2$, of about $0.5$ independent of $x_L$ .  

From the theoretical point of view, it would be best to observe leading neutrons produced in DIS at very large $Q^2$ where the rescattering absorptive corrections are negligible; and to measure $F_2^{\pi}$ in a most direct and clear way.  Unfortunately, the event rate at large $Q^2$ is limited.  The ZEUS preliminary data \cite{dis06} correspond to $Q^2 > 2~{\rm GeV}^2$, with an average, $\langle Q^2 \rangle$, of 16 ${\rm GeV}^2$, so we cannot neglect absorption even in the DIS data sample.  To be precise we have to integrate over the size of the $\qq$ pair produced by the photon, starting from the hadronic/confinement scale $\sim \Lambda_{\rm QCD}^2 \sim 0.1~~{\rm GeV}^2$ up to $\langle Q^2 \rangle$.  The part of the cross section originating from a small size $\qq$ pair will have negligible absorption, while a large size pair will be strongly affected by rescattering and will suffer an $S^2$ suppression. This prescription was implemented explicitly in Refs.~\cite{ap,gw}. Here we adopt a simplified approach assuming that the part of $F_2^\pi$ measured at the initial hadron scale\footnote{We take the scale to be $m^2_{\rho}=0.6~ {\rm GeV}^2$.} enters with the same absorptive factor as in photoproduction, while the remainder of $F_2^\pi$, which is generated by DGLAP evolution with strong $k_t$-ordering, that is by a small size $\qq$ pair, is taken to have $S^2=1$.

\begin{figure}
\begin{center}
\includegraphics[height=15cm]{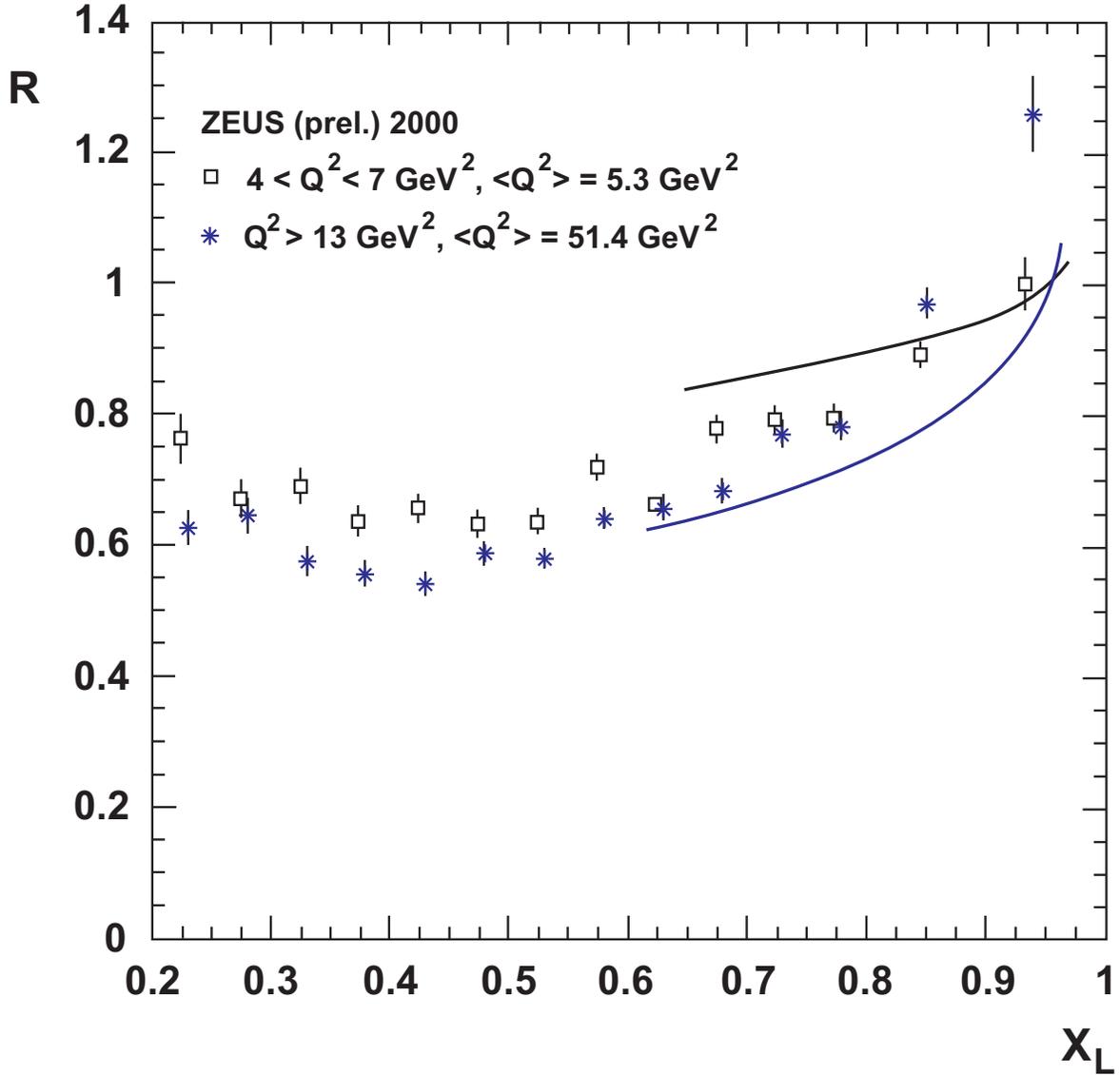}
\caption{The ratio, $R=$ photoprod./DIS, of leading neutrons from photoproduction, $(d\sigma(Q^2 \simeq 0)/dx_L)/\sigma(0)$, and DIS production, $(d\sigma(Q^2)/dx_L)/\sigma(Q^2)$ as a function of $x_L$ for two different intervals of $Q^2$, namely $4<Q^2<7~{\rm GeV}^2$ and $Q^2>13~{\rm GeV}^2$, compared with predictions for $Q^2=5.3~{\rm GeV}^2$
(upper curve for $x_L<0.9$) and $Q^2=51.4~{\rm GeV}^2$ (lower curve). The predictions include the $\rho$ and $a_2$ exchange contributions, as well as $\pi$-exchange, and the rescaled value of $Q^2$ (eqs. (\ref{DISmod}), (\ref{aqmphmod})), as described in Section 4. The points are preliminary ZEUS data \cite{blois}. }
\label{fig:3}
\end{center}
\end{figure}

Since the absorptive effects in the present DIS data are not negligible, the ratio of the probabilities of observing leading neutrons in photoproduction compared to DIS production will be closer to 1 than the ratio of the curves shown with and without absorption in Fig.~\ref{fig:2}.  The ratio is plotted in Fig.~\ref{fig:3} for two different values of $Q^2$.  The predictions are calculated with the same cuts as the ZEUS leading neutron data \cite{blois}.  The growth of the ratio as $x_L \to 1$ reflects the fact that the energy (that is, the $x$) dependence of DIS production, $\sigma \sim x^{-\lambda}$, is much steeper than for photoproduction. Since $x$ in the $\gamma \pi$ interaction is smaller by a factor $(1-x_L)$ than in the proton structure function used to normalise the neutron yields, $(d\sigma /dx_L)/\sigma$, this gives an extra factor
\be
(1-x_L)^{\lambda(0)-\lambda (Q^2 \neq 0)}
\ee
in the ratio plotted\footnote{We use MRST2001 LO partons \cite{mrstlo}.} in Fig.~\ref{fig:3}.  For $x_L \sim 0.7$, where this factor is not so important, the ratio $R$ decreases with increasing $Q^2$ since at larger $Q^2$ we have smaller absorption. Of course the prediction of the ratio depends on the value chosen for $\sigma_{\rm abs}$ and the ansatz used for $F_2^\pi$. For example, if we used the rescaled value of $Q^2$ in eq. (\ref{DIS}) then we would increase the predicted ratio by $10-20\%$. Turning the argument around, we see that if the ratio is precisely measured for different $Q^2$ values, then it would be possible to determine both $F_2^\pi$ and $\sigma_{\rm abs}$.  

\section{The $p_t$ dependence of leading neutrons}
\begin{figure}
\begin{center}
\includegraphics[height=15cm]{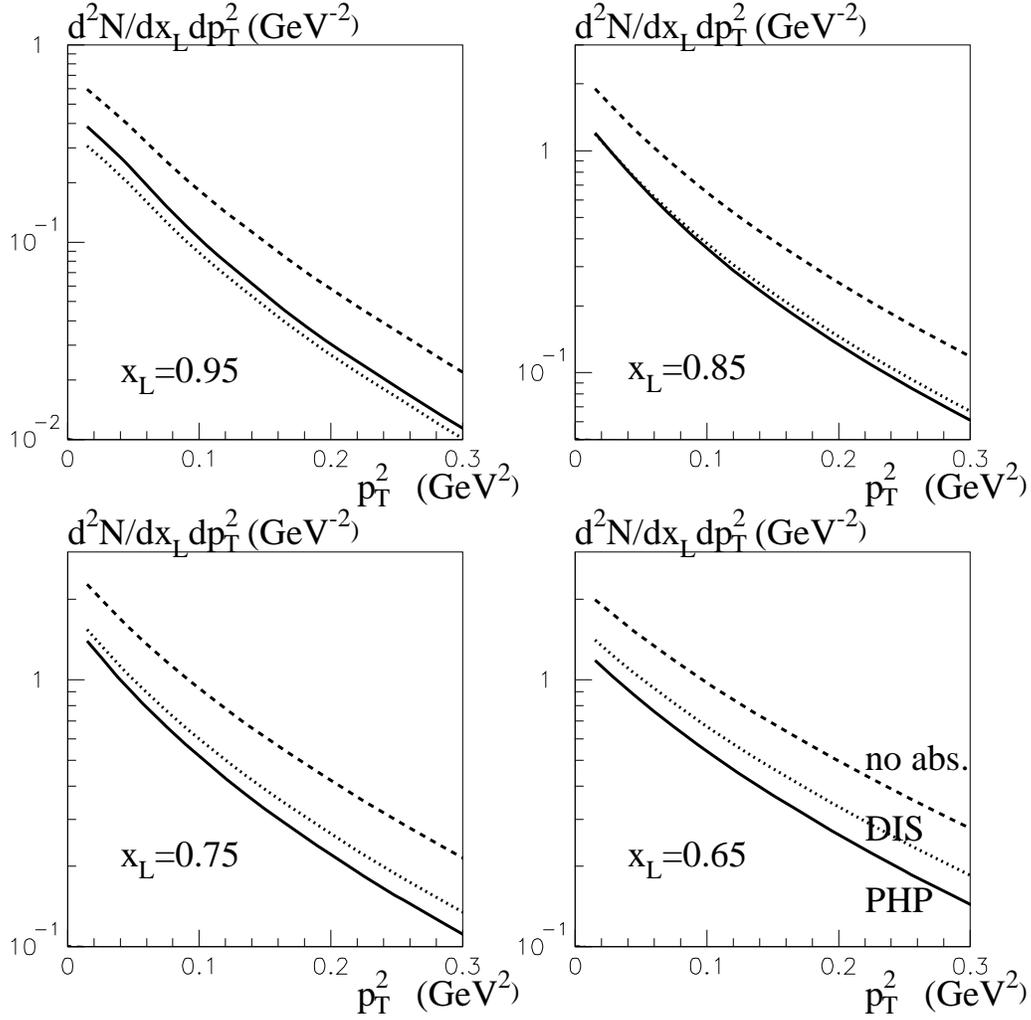}
\caption{The $p_t^2$ distributions of leading neutrons for four different values of $x_L$. The lower two curves correspond to photoproduction (PHP) and production by deep inelastic scattering (DIS) with $Q^2 = 16~{\rm GeV}^2$. The upper curve is obtained assuming that there is no absorption.}
\label{fig:4}
\end{center}
\end{figure}

\begin{figure}
\begin{center}
\includegraphics[height=15cm]{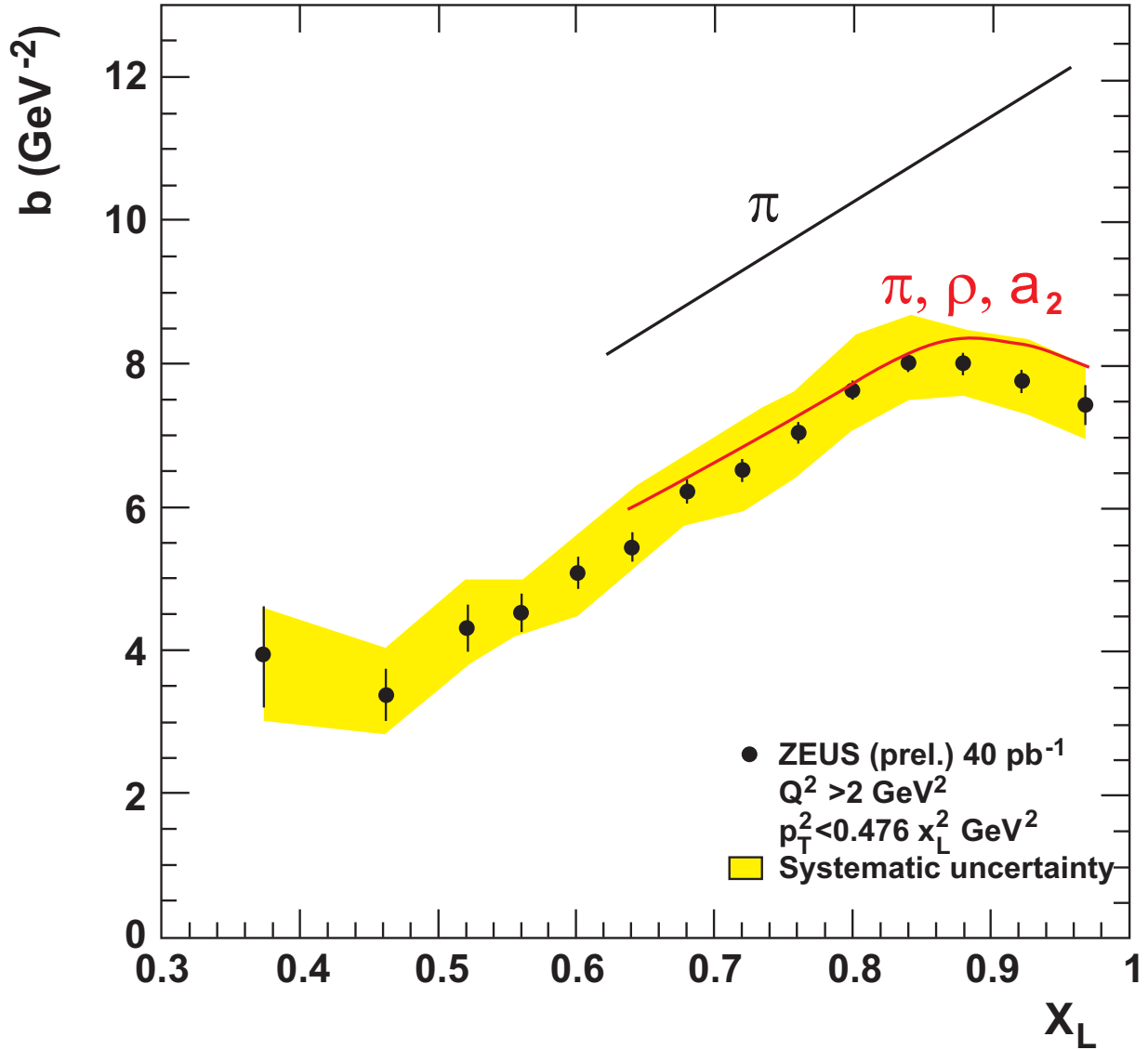}
\caption{The slope, $b$, of the $p^2_t$ distribution of leading neutrons produced in DIS as a function of $x_L$. The data are taken from \cite{dis06}. The upper curve corresponds to $\pi$ exchange, with absorption and migration effects included as in Ref.~\cite{KKMRln}, but with the same kinematic cuts as used to obtain the data \cite{dis06}. The lower curve is calculated including $\rho$ and $a_2$ exchanges, in addition to $\pi$ exchange, as described in Section 4. The decrease of the slope for $x_L \gapproxeq 0.85$ reflects the vanishing of $t_{\rm min}$ as $x_L \to 1$.}
\label{fig:5}
\end{center}
\end{figure}

\begin{figure}
\begin{center}
\includegraphics[height=12cm]{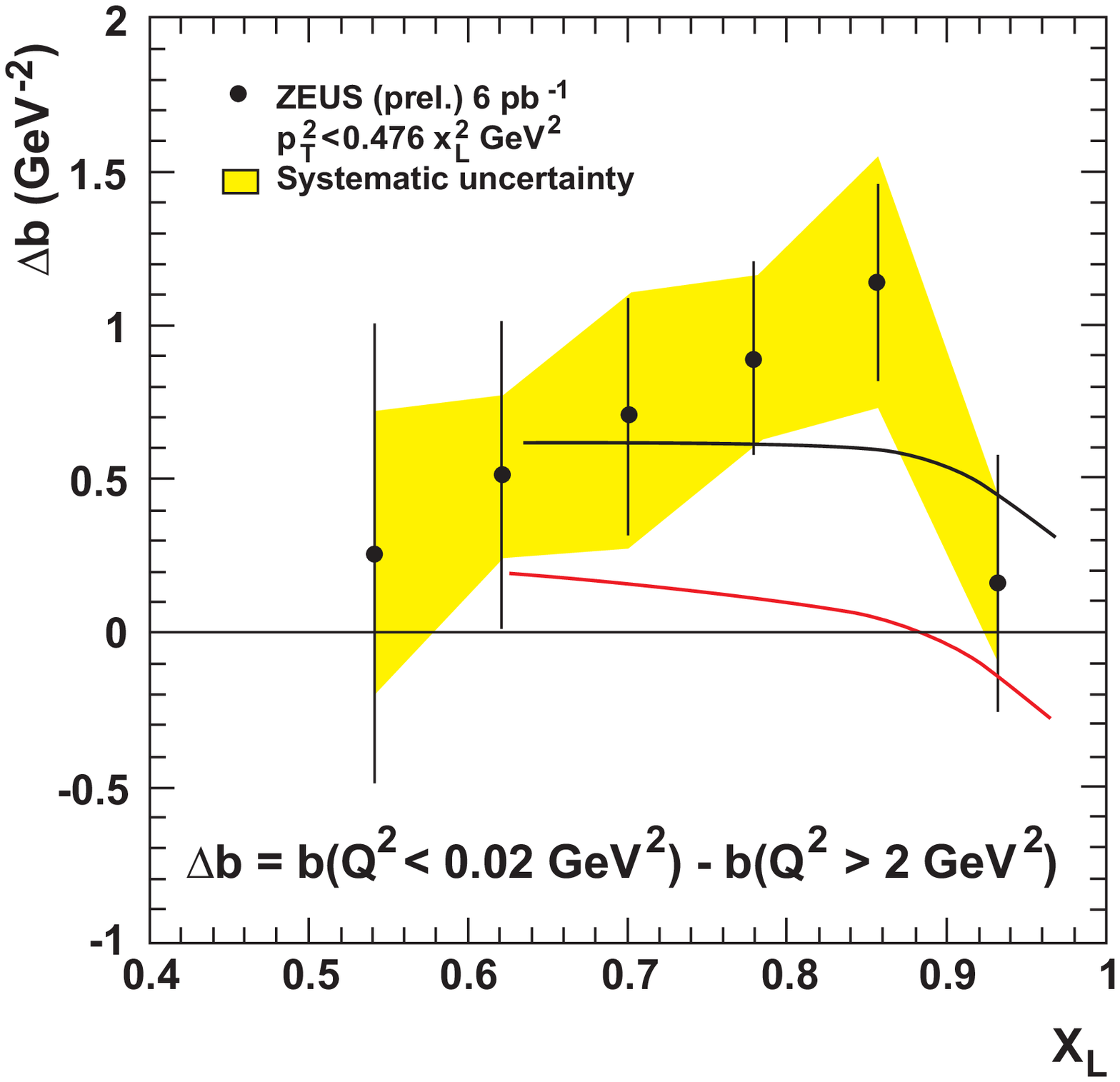}
\caption{The difference in the slopes of the $p_t$-distributions of leading neutrons coming from photoproduction and from deep inelastic scattering. We compare the preliminary ZEUS data \cite{dis06} with the calculations based on $\pi$ exchange (upper curve) and then including also $\rho$ and $a_2$ exchange (lower curve).}
\label{fig:5a}
\end{center}
\end{figure}

The predicted $p_t$ dependence of leading neutrons is shown in Fig.~\ref{fig:4}, for different values of $x_L$. We see that the distributions do not have exactly exponential form, as may be anticipated from (\ref{pi}). However the departure from this form is not large, and the experimental data are usually discussed in terms of an average slope $b$, where $d\sigma/dp^2_t \sim {\rm exp}(-bp^2_t)$.  Note that, in spite of the factor of $t$ in the numerator of (\ref{pi}), the distributions do not vanish at $p_t=0$, due to
\be
t_{\rm min}~\simeq~-(1-x_L)^2 m_N^2/x_L;
\ee
where $t=-(p^2_t/x_L)+t_{\rm min}$.  The calculated distributions qualitatively agree with the data.  The slope, $b$, grows with $x_L$, as can been seen in Fig.~\ref{fig:5}. This arises from the factor $(1-x_L)^{-2\alpha_\pi (t)}$ in (\ref{pi}), and reflects the Reggeization of the pion.  The predicted slope for photoproduction of leading neutrons is a bit larger than for DIS production, see Fig.~\ref{fig:5a}, since strong absorption at small impact parameters, $\rho_T$, effectively pushes the distribution into the larger $\rho_T$ region.

Unfortunately, Fig.~\ref{fig:5} shows that the predicted $p_t$ distribution is too steep in comparison with the data. This is a general property of any model, see \cite{dis06}.  Of course, there is a possibility to modify eq. (\ref{pi}), using a more complicated signature factor. If we replace the pion propagator, $1/(m_\pi^2-t)$, by the usual Reggeon signature factor, $\pi \alpha_\pi^\prime /(2~{\rm sin}(\pi \alpha_\pi (t)/2))$, then the slope is a bit smaller, but only by $0.2-0.4 ~{\rm GeV}^{-2}$. In some studies, motivated by Veneziano-type models, the signature factor is written in terms of the $\Gamma$ function. Such a function, $\Gamma (-\alpha_\pi (t))$, grows at large negative $t$, but, first, it does not help in the region of interest, $-t<0.5~{\rm GeV}^2$, and, second, this growth reflects the contribution of heavier mesons which lie on daughter trajectories. The simplest possibility, to reconcile the calculated slopes with the measure values, is to take a negative slope of about 1 ${\rm GeV}^{-2}$ in the form factor $F(t)$ in (\ref{pi}), but this is physically unacceptable.

\section{Including $\rho$ and $a_2$ exchange contributions}

This persistent discrepancy between the measured and predicted $p_t$-slopes of leading neutrons from $\pi$-exchange leads us to investigate the contributions of secondary $(\rho, a_2)$ Reggeon exchanges. Note that the spin-flip $a_2$-exchange amplitude interfers with the $\pi$-exchange amplitude. These contributions were discussed in Ref.~\cite{kop}, where it was stated that they never exceed $10\%$ of the $\pi$-exchange induced cross section. But, there, the authors concentrated on very low $p_t$, to be close to the $\pi$ pole, and did not allow for the spin-flip $\rho$ and $a_2$ nucleon-Reggeon vertex. The exchange of the $\rho$-trajectory (but not of the $a_2$-trajectory) was
considered also in \cite{Nrho}. There the $\rho$
contribution was estimated to be about 20$\%$ of the whole cross
section for events with a leading neutron with $x_L=0.7 -
0.8$. As a result the authors pointed out that including $\rho$ exchange will lead to a
smaller slope in the $p_t$ distribution of leading neutrons.  

For $\rho$-exchange, spin flip is very large \cite{iw}, 
\be
\frac{V_\rho^{\rm flip}}{V_\rho^{\rm non-flip}}~\simeq~ 8~\frac{p_t}{2m_N}.
\ee
To evaluate the effect of $\rho,~a_2$ exchange, we use the parametrization of Ref.~\cite{iw} for the $\rho$ meson, assume $\rho -a_2$ exchange degeneracy, and assume the same additive quark model relations for the $\gamma \rho$ and $\gamma a_2$ cross sections, which we used for $\gamma \pi$, that is (\ref{aqmph}) and (\ref{DIS}). Clearly these extra exchange contributions will enlarge the cross section for leading neutron production, mainly at the larger $p_t$ values, and less as $p_t \to 0$, which will lead to a smaller average slope.  Indeed the calculated slope is in more agreement with the data, but now the probability to observe leading neutrons becomes too large. That is by using the same values of the parameters that we took in \cite{KKMRln}, and the additive quark model relations of (\ref{aqmph}) and (\ref{DIS}), the calculated cross section is nearly $40\%$ above the data.  

So we have better agreement with the data for the slope $b$, at the expense of now failing to describe the size of the cross section. Can we do better? Of course, there are many parameters which are not precisely known.  In an attempt to achieve a simultaneous description of the cross section and the slope of the $p_t^2$ distribution of leading neutron production, we change the parameters within the limits of acceptability.  We take the slope of the pion trajectory to be $\alpha_{\pi}^\prime = 0.9$ rather than $1~ {\rm GeV}^{-2}$. We use the normal Reggeon signature factor of the pion $\pi \alpha_\pi^\prime /(2~{\rm sin}(\pi \alpha_\pi (t)/2))$. We increase the absorption by setting $C=1.6$ and $\sigma_{\rm tot} (\pi p) =34$ mb,
rather than\footnote{There is evidence that the factor $C$, which accounts for the diffractive excitation, should be larger \cite{abk}. At first sight we might expect that stronger absorption would be generated by including the contribution from the ``enhanced'' diagrams, which arise from rescattering from intermediate partons. However then there would be a strong energy dependence for the probability to produce leading neutrons. This is not observed in the data. Further discussion is given in Ref.~\cite{KKMRln}.} $C=1.3$ and $\sigma_{\rm tot} (\pi p) =31$ mb as in \cite{KKMRln}.  This has the effect of reducing\footnote{This is more in line with an earlier calculation \cite{KKMR034}
which gave a rapidity gap survival factor $S^2$ of 0.34 for the resolved part
of the photon wave function.} $S^2$ from 0.5 to 0.4. In addition we should be more careful with the form of the additive quark model relations, (\ref{aqmph}) and (\ref{DIS}), for the $\gamma$-meson cross sections. Let us discuss the perturbative QCD expectations. At lowest order in $\alpha_S$, the cross sections are proportional to the radius squared of the hadron, $\sigma \sim \alpha_S^2 r^2$ \cite{low,nuss,brod,kopl}.  If the two colliding particles have quite different radii, $r_a \ll r_b$, the cross section is controlled by the smallest radius. Thus in DIS, the $\gamma$-meson cross section $\sigma \sim 1/Q^2$, as in (\ref{sigF}). However the interval of evolution of the structure function $F_2$ starts from the largest radius $r_b$. Thus we should rescale the value of $Q^2$ in (\ref{DIS}), and take
\be
F_2^{\rm meson} (x,Q^2)~=~\frac{2}{3}~F_2^p \left(\frac{2}{3}x,\frac{r^2_m}{r^2_p}Q^2\right).
\label{DISmod}
\ee
Moreover, we assume that the quark wave functions in the $\rho$ and $\pi$ mesons are the same (as would follow from SU(6) symmetry).
We take $r^2_m = 0.44 ~{\rm fm}^2$ \cite{rpi} and $r^2_p = 0.76 ~{\rm fm}^2$ \cite{rp}. On the other hand, for photoproduction, where the two radii are comparable, it is reasonable to replace $r^2$ by $r_m r_p$. Then we have
\be
\sigma^{\rm tot}_{\gamma-{\rm meson}}~=~\frac{2}{3}~\frac{r_m}{r_p}~\sigma^{\rm tot}_{\gamma p},
\label{aqmphmod}
\ee
in the place of (\ref{aqmph}).

\begin{figure}
\begin{center}
\includegraphics[height=15cm]{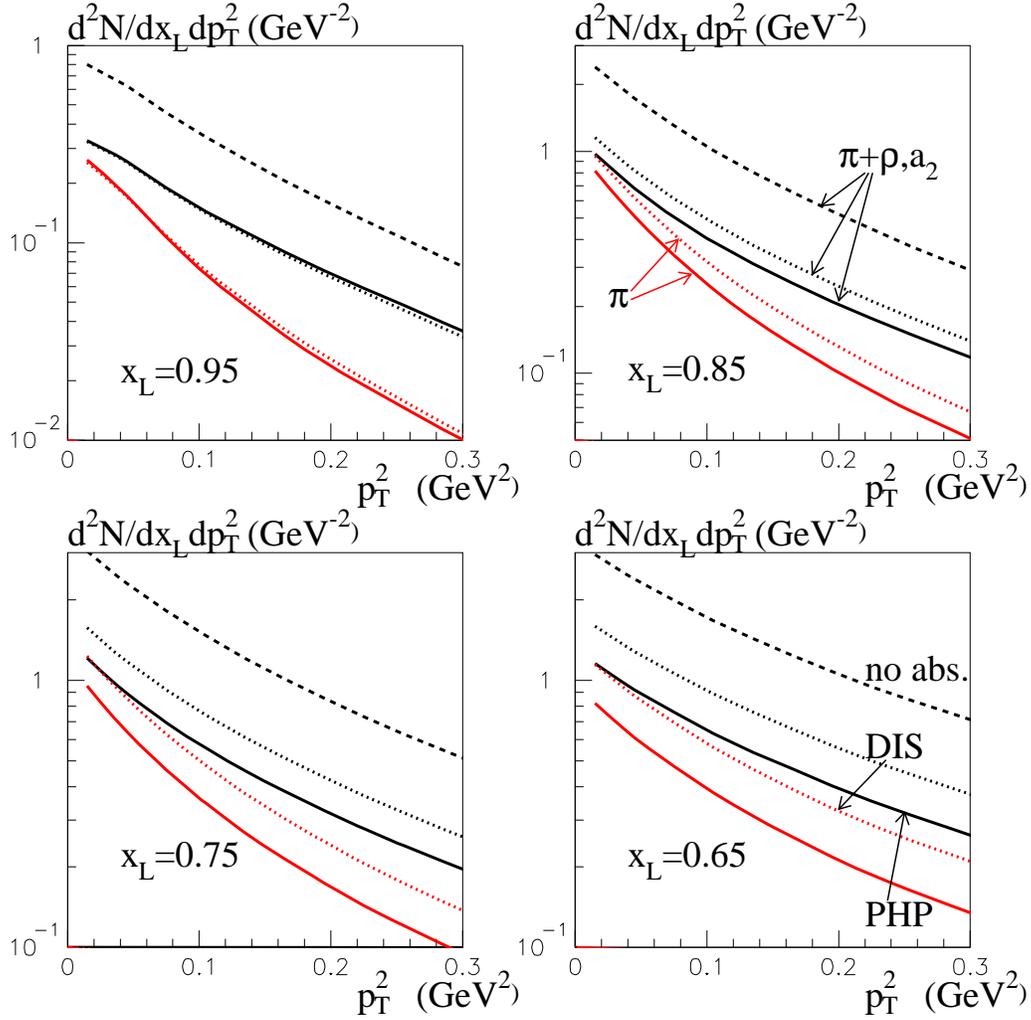}
\caption{The $p_t^2$ distributions of leading neutrons for four different values of $x_L$ as in Fig.~\ref{fig:4}, but now including the $\rho$ and $a_2$ exchange contributions, with increased absorption, as described in the text. For completeness, we also show, by the two lower curves, the individual $\pi$ exchange contribution for DIS (dotted) and photoproduction (continuous) using the modified prescription.}
\label{fig:6}
\end{center}
\end{figure}

The predictions obtained with these modifications are shown in Fig.~\ref{fig:2} and Fig.~\ref{fig:5}. Indeed we do achieve a more satisfactory simultaneous description of the cross section and the $p_t^2$-slopes as functions of $x_L$. The results of Fig.~\ref{fig:4} now become those shown in Fig.~\ref{fig:6}. The lower pair of curves correspond to just the $\pi$-exchange contribution. By comparing the $\pi$ component with the full contribution, we see that, for very low $p_t$, $\pi$-exchange provides more than $70\%$ of the cross section for $x_L>0.7$, whereas at $p^2_t \sim 0.3~ {\rm GeV}^2$ the $(\rho,~ a_2)$-exchange contribution starts to dominate. Finally, the curves plotted in Fig.~\ref{fig:3} correspond to the modified description.

\section{Conclusions}

Leading neutron spectra at HERA have a rich structure. The data offer a serious challenge; it is difficult to describe the dependence of the cross section on $x_L$ and $Q^2$, while at the same time satisfactorily reproducing the observed $p_t$ behaviour.  First pure Reggeized $\pi$-exchange fails on all counts.  The cross section data indicate the important role of absorptive corrections, arising from rescattering.  This effect suppresses the yield of leading neutrons by a factor of about 0.5 in photoproduction, but less in DIS production.  For the same reason the slope of the $p_t$ distribution in photoproduction is a little larger that in DIS production. The calculation, based on Ref.~\cite{KKMRln}, but updated to account for the new experimental cuts gives a satisfactory description of the observed cross section for $x_L \gapproxeq 0.6$. 

However, although the difference in slopes for photoproduction and DIS production is reasonable, the predicted $p_t$ distribution for photoproduction falls off more quickly than the data. This is a long standing general problem for $\pi$-exchange models.  Indeed, the relatively low experimental value of the slope $b$ indicates the presence of secondary $(\rho,~ a_2)$ Regge spin-flip contributions. It is desirable to measure the slope in different intervals of $p_t$ since it offers the chance to separate $\pi$ from $\rho$ and $a_2$ exchanges. Recall that for $\pi$ exchange, the vanishing of the amplitude as $t \to 0$ is almost compensated by the $\pi$ propagator $1/(t-m^2_\pi)$. On the other hand spin-flip $\rho$-exchange is indeed proportional to $t$. Therefore in the largest $x_L$ interval, where $t_{\rm min}$ becomes small enough the presence of a significant $\rho$ contribution would produce a ``kinematic turnover'' in the slope. In fact, neglecting $t_{\rm min}$, the spin-flip $\rho$ contribution leads to a negative slope as $p_t \to 0$. This behaviour explains the different behaviour of the two curves shown in Fig.~\ref{fig:5} as $x_L \to 1$.

\begin{figure}
\begin{center}
\includegraphics[height=15cm]{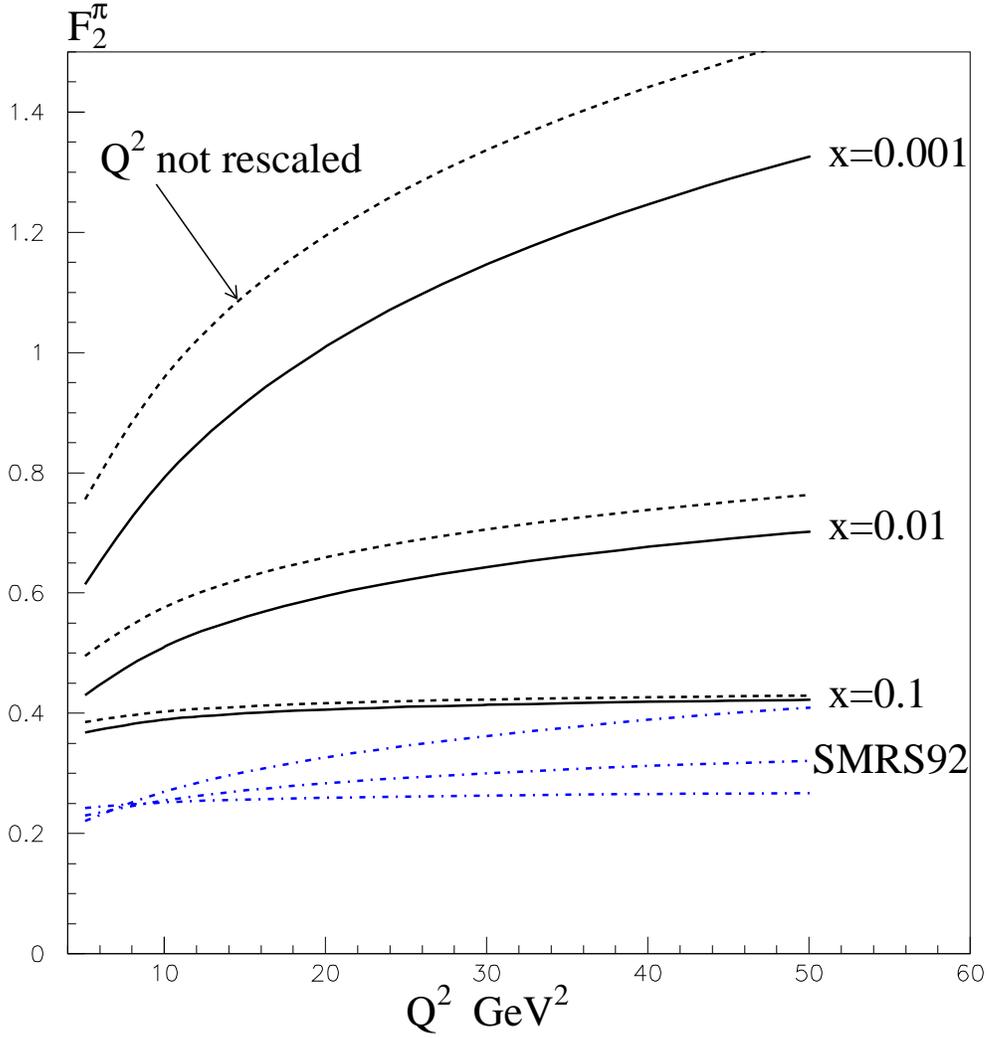}
\caption{The pion structure function, $F_2^{\pi}(x,Q^2)$, as a function of $Q^2$ for three different values of $x$. We also show the structure function calculated from the partons of \cite{SMRS}. On this plot $x$ denotes the Bjorken-$x$ of the pion, $x_{\pi}$, that is $x_{\pi}=x/(1-x_L)$.}
\label{fig:8}
\end{center}
\end{figure}

Although the presence of the secondary trajectories improve the description of the absolute slope, they disturb the satisfactory description of the cross section obtained by just $\pi$ exchange.  We were required to look in more detail at the calculation of the survival factor. A physically-motivated recalculation of absorption yielded $S^2=0.4$. In this way we achieved a more satisfactory overall description of the leading neutron spectra. The original and the modified pion structure functions, $F_2^{\pi}(x,Q^2)$, that is (\ref{DIS}) and (\ref{DISmod}), are plotted in Fig.~\ref{fig:8} as functions of $Q^2$ for three different $x$ values, together with predictions obtained from parton distributions of the pion. 

Already an initial analysis of the preliminary ZEUS data have yielded interesting results. To obtain a reasonable description of the all features of the leading neutron data we are lead to select the pion structure function given by the continuous curves in Fig.~\ref{fig:8} in preference to those given by the dashed curves. Moreover, the pion structure function preferred by the data is found to be larger than that obtained using the parton distributions of the pion, see, for example, \cite{SMRS}.  Here, we should note that the parton distributions of the pion were obtained from an analysis of high statistics data, from $\pi^\pm N$ experiments for both Drell-Yan and prompt photon production, which did not extend below $x_{\pi} \sim 0.2$.  So extrapolation is necessary to obtain the SMRS curves in Fig.~\ref{fig:8}.  

In summary, the physically-motivated modification, (\ref{DISmod}), while not perfectly reproducing every feature of the leading neutron data, is clearly a step in the right direction. The analysis described here should be regarded as a broad-brush exploratory study. In our additive quark model type of approach we have forced the structure function of the pion to mimic that of the proton.  There is no reason why it should, and so some mismatch with the data is to be expected. Indeed we can expect a more detailed analysis of more precise data with a freely parametrised pion structure function, $F_2^\pi (x,Q^2)$, will give a quantitative measure of the behaviour of the function, as well as the properties of the rapidity gap survival factor, $S^2$, that is, of absorptive effects.    

\section*{Acknowledgements}

We thank Aliosha Kaidalov, Jon Butterworth, Bill Schmidke and Michele Arneodo for valuable discussions. MGR would like to thank the IPPP at the University of Durham for hospitality, and ADM thanks the Leverhulme Trust for an Emeritus Fellowship. This work was supported by the Royal Society,
the UK Particle Physics and Astronomy Research Council, by grant RFBR 04-02-16073 and by the Federal Program of the Russian Ministry of Industry, Science and Technology SS-1124.2003.2.

\section*{Appendix}
It is simplest to calculate the absorptive corrections in impact parameter, $\rho_T$, space. For convenience, we adopt the notation used in Ref.~\cite{KKMRln}. We take the Fourier transform of the original `bare' amplitude, $A$, which corresponds to the cross section given in ({\ref{pi}) without the $S^2$ factor. It depends on the impact parameter $\vec{\rho}_{\pi n}$. The subscript $\pi$ denotes
the `transverse' position of the $\pi-\pi-$Pomeron vertex in Fig.~\ref{fig:1}(b). That is, $\rho_{\pi n}$
is Fourier conjugate of the neutron transverse momentum in the
bare $\pi$-exchange amplitude. Then the transverse distance between
the leading baryon and the incoming photon (or $q\bar q$-pair) is
\be
\rho_T \equiv \rho_{\gamma n}~=~|\vec{\rho}_{\pi n}+\vec{\rho}_{\gamma\pi}|,
\ee
where $\rho_{\gamma\pi}$ is the
impact parameter for the amplitude describing the interaction of the incoming photon
with the `effective' pion (exchanged in the $t$-channel).  For a fixed value of $\vec{\rho}_{\gamma\pi}$, the suppression of amplitude $A$ is given by the eikonal exp$(-\Omega/2)$, with
\be
\label{m3}
\Omega(\rho_T)\ =\ \frac{\sigma_{\rm abs}}{4\pi B} ~{\rm exp}(-\rho^2_T/4B)\ ,
\ee
where the slope $B$ of the $\rho\pi$ amplitude is taken to be 5 ${\rm GeV}^{-2}$. Thus to determine $p_t$ we must multiply $A(\vec{\rho}_{\pi n})$ by the eikonal factor  exp$(-\Omega/2)$, and perform the inverse Fourier transform
\be
A(x_L,\vec{p}_t;\vec\rho_{\gamma\pi})~=~\int \frac{d^2 \rho_{\pi n}}{2\pi}~e^{i\vec p_t\cdot \vec\rho_{\pi n}}~A(x_L,\vec{\rho}_{\pi n})~e^{-\Omega(\rho_T)/2}.
\ee
Finally, we integrate over $\vec{\rho}_{\gamma\pi}$ to obtain the cross section
\be
\frac{d\sigma}{dx_Ldp^2_t}~=~\int d^2\rho_{\gamma\pi}~\frac{|A(x_L,\vec p_t;\vec\rho_{\gamma\pi})|^2}{4\pi B}~e^{-\rho^2_{\gamma\pi}/4B},
\ee
suppressed by rescattering effects.



\begin{thebibliography}{x}

\bibitem {KKMRln}A.B.~Kaidalov, V.A.~Khoze, A.D.~Martin and M.G.~Ryskin,
{\tt arXiv:hep-ph/0602215}, Eur.\ Phys.\ J. {\bf C} in press.

\bibitem {KMRsoft} V.A.~Khoze, A.D.~Martin and M.G.~Ryskin,
Eur.\ Phys.\ J. {\bf C18} (2000) 167.

\bibitem{GpiN} V. Stoks, R. Timmermans and J.J. de Swart, Phys. Rev.
{\bf C47} (1993) 512;\\
R.A. Arndt, I.I. Strakovsky, R.L. Workman and M.M. Pavan, Phys. Rev.
{\bf C52} (1995) 2120.

\bibitem{dis06} ZEUS collaboration: presented by M. Soares at the XIV conf. on Deep Inelastic Scattering, DIS 2006, Tsukuba, Japan, April 2006.

\bibitem{blois} ZEUS collaboration: presented by M. Soares at the 11th Int. Conf. on Elastic and Diffractive Scattering, Blois, France, May 2005.

\bibitem{ap} U.D. Alesio and H.J. Pirner, Eur. Phys. J. {\bf A7} (2000) 109.

\bibitem{gw} K.J. Golec Biernat and M. Wusthoff, Phys. Rev. {\bf D59} (1999) 014017; {\bf D60} (1999) 114023.

\bibitem {mrstlo} A.D. Martin, R.G. Roberts, W.J. Stirling and R.S. Thorne, Phys. Lett. {\bf B531} (2002) 216.

\bibitem{kop} B.Z. Kopeliovich, B. Povh and I. Potashnikova, Z. Phys. {\bf C73} (1996) 125.

\bibitem{Nrho} H. Holtmann, G. Levman, N.N. Nikolaev, A. Szczurek and J. Speth, Phys. Lett. {\bf B338} (1995) 393.

\bibitem{iw} A.C. Irving and R.P. Worden, Phys. Rept. {\bf 34} (1977) 117.

\bibitem{abk} A.B. Kaidalov, Phys. Rept. {\bf 50} (1979) 157; and private communication.

\bibitem {KKMR034} A.B. Kaidalov, V.A. Khoze, A.D. Martin and M.G. Ryskin, Phys. Lett. {\bf B567} (2003) 61.

\bibitem {low} F.E. Low, Phys. Rev. {\bf D12} (1975) 163.

\bibitem {nuss} S. Nussinov, Phys. Rev. Lett. {\bf 34} (1976) 1286.

\bibitem {brod} G. Bertsch, S.J. Brodsky,  A.S. Goldhaber  and J.G. Gunion
 Phys. Rev. Lett. {\bf 47} (1981) 297.

\bibitem {kopl} B.Z. Kopeliovich, L.I. Lapidus and A.B. Zamolodchikov,
JETP Lett. {\bf 33} (1981) 595 [Pisma Zh.Eksp.Teor.Fiz. {\bf 33} (1981) 612].

\bibitem {rpi} S.R. Ameldolia et. al., Nucl. Phys. {\bf B227} (1986) 168. 

\bibitem {rp} R. Rosenfelder et al., Phys. Lett. {\bf B479} (2000) 381. 

\bibitem {SMRS} P.J. Sutton, A.D. Martin, R.G. Roberts and W.J. Stirling, Phys. Rev. {\bf D45} (1992) 2349.
\end{thebibliography}
\end{document}